\documentclass[prl,aps,amssymb,twocolumn,showpacs,preprintnumbers,superscriptaddress]{revtex4}
\usepackage{graphicx}
\begin{document}

\title{Chaos-assisted nonresonant optical pumping of quadrupole-deformed microlasers}


\author{Sang-Bum Lee}
\address{School of Physics and Astronomy, Seoul National University, Seoul 151-747, Korea}

\author{Jeong-Bo Shim}
\address{Department of Physics, Korea Advanced Institute of Science and Technology, Taejon 305-701, Korea}

\author{Sang Wook Kim}
\address{Department of Physics Education, Pusan National University, Pusan 609-735, Korea}

\author{Juhee Yang}
\address{School of Physics and Astronomy, Seoul National University, Seoul 151-747, Korea}

\author{Songky Moon}
\address{School of Physics and Astronomy, Seoul National University, Seoul 151-747, Korea}

\author{Hai-Woong Lee}
\address{Department of Physics, Korea Advanced Institute of Science and Technology, Taejon 305-701, Korea}

\author{Jai-Hyung Lee}
\address{School of Physics and Astronomy, Seoul National University, Seoul 151-747, Korea}

\author{Kyungwon An}
\email{kwan@phya.snu.ac.kr}
\address{School of Physics and Astronomy, Seoul National University, Seoul 151-747, Korea}
\date{\today}

\begin{abstract}
Efficient nonresonant optical pumping of a high-$Q$ scar mode in a two-dimensional quadrupole-deformed microlaser has been demonstrated based on ray and wave chaos. Three-fold enhancement in the lasing power was achieved at a properly chosen pumping angle. The experimental result is consistent with ray tracing and wave overlap integral calculations.
\end{abstract}


\maketitle


Optical pumping is frequently used for microcavity lasers for its easy implementation and high efficiency. However, in some cases it is not very effective, particularly for circular microcavities. This is because an optical beam or a ray refracted in a circular cavity immediately exits the cavity by subsequent refraction due to the cavity symmetry and therefore the low-order whispering gallery modes (WGM's) which are distributed near the cavity boundary are not well excited. 

\begin{figure}
\includegraphics[width=3.3in]{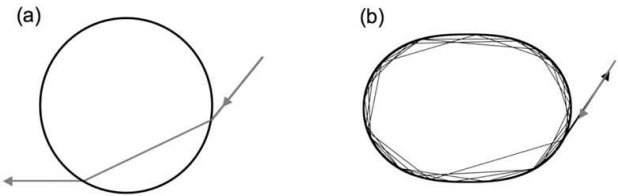}
\caption{Comparison of nonresonant pumping schemes in ray picture for (a) a circular cavity  and (b) a quadrupole-deformed microcavity.}
\label{figure1}
\end{figure}

The conventional resonant pumping based on evanescent-wave coupling by using a prism \cite{Sandoghdar96} or a tapered fiber \cite{vahala02} can increase the pumping efficiency by many orders of magnitude. However, the resonant pumping requires special arrangements for optimal coupling and fine wavelength tuning of the pump to a particular cavity mode, and thus may not be practical for many applications.

Asymmetrical resonant cavities (ARC's) \cite{Nockel97,Gmachl98}, well known for chaotic ray dynamics inside, have an advantage in this regard. They can also support high $Q$ modes distributed near the cavity boundary similar to the WGM. Due to the asymmetry imposed by the cavity boundary, a ray can be injected in and circulate near the cavity boundary for many round trips. This can be understood by considering its time reversed process. When a ray is launched inside an ARC at an angle larger than the critical angle $\chi_c = \sin^{-1}(1/n)$ with $n$ the index of refraction of the cavity medium, the ray can undergo total internal reflections until subsequent incident angles would become larger than $\chi_c$ due to the cavity asymmetry and thus the ray would refract out. If we trace the ray trajectory in reverse, we then find a ray which would perform efficient optical pumping for high-$Q$ modes \cite{Lee02} distributed near the cavity boundary. 

In practice, the pump beam width is usually much larger than the cavity width and thus a wide pumping, covering the entire width of the cavity, is performed. Nonetheless, it is still expected that a significant portion of the pump beam can be efficiently coupled into the cavity if the pump beam is incident on the ARC at a properly chosen angle and therefore its intensity can be significantly built up inside. 

\begin{figure}
\includegraphics[width=3.3in]{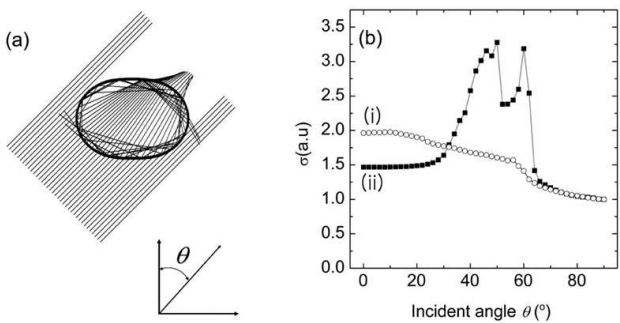}
\caption{(a) Ray model for the calculation of the intracavity pump intensity when pumped by a plane wave. (b) Average density of ray as function of the angle of incidence for (i) elliptical and (ii) quadrupolar microcavities. }
\label{Raymodel}
\end{figure}

The efficiency of the non-resonant wide pumping of a ARC from the side can be estimated by a ray tracing analysis. We consider a particular type of ARC in the present study, namely a quadrupole-deformed microcavity (QDM), the boundary of which is defined in the polar coordinates by the relation $\rho(\phi)=a(1+\eta \cos 2\phi)$, where $a$ is a mean radius and $\eta$ is a deformation parameter. We treat the incident pump beam as a set of a large number ($\sim 10^4$) of equally-spaced parallel rays as sketched in Fig.\ \ref{Raymodel}(a). For a given incident angle $\theta$, measured with respect to the minor axis of QDM, we then follow each ray (let us say the $i$th ray) until it refracts out the cavity and then calculate the distance $l_i(\theta)$ it has traveled inside before the refractive escape. Then the average density of ray in the cavity is defined as $\sigma(\theta)=\sum_i l_i (\theta)/\sum_i l_i (\pi/2)$, normalized with respect to $\theta=\pi/2$.

This ray model, however, is too much simplified to be used for a realistic situation. We must take into account the transmission coefficient into the cavity and the reflection coefficient upon each subsequent reflection, not to mention the absorption of the cavity medium. The length $l_i$ is then interpreted as an effective length to be given by
\begin{equation}
l_i=\sum_{n=0}^\infty \left[\prod_{j=0}^n R_{ij}\int_{s_n^{(i)}}^{s_{n+1}^{(i)}}P(r){\rm e}^{-\alpha s}ds\right]\;,
\end{equation}
where $R_{i0}$ is the transmission coefficient into the cavity for the $i$th ray, $R_{ij}$ ($j>0$) is the reflection coefficient upon the $j$th reflection of the $i$th ray off the cavity boundary, $s_n^{(i)}$ is the ray path length between the $n$th and the $(n+1)$th reflection, $\alpha$ is the absorption coefficient of the cavity medium and $r$ is the radial coordinate of the ray. The function $P(r)$ is unity for $\rho/n<r<\rho$ and zero for $r<\rho/n$ with $\rho$ the radial coordinate of the cavity boundary as defined above. The purpose of $P(r)$ is to include only the ray segment lying in the limited region that a ray undergoing total internal reflection would occupy.

The result of calculation is shown in Fig. \ref{Raymodel}(b) for a QDM with $\eta$=16\%, $\alpha$=1.65 cm$^{-1}$, $a$=15 $\mu$m and $n=1.361$, which are the parameters corresponding to the experiment to be discussed below. It is found that the average density of ray is maximized at $\theta\simeq 50^\circ$, which means that rays launched above critical angle inside the cavity would refract out at $\theta\simeq 50^\circ$ after many reflections. 
We also calculated the average ray density for an elliptical-deformed cavity with a similar set of parameters. Since the ray motion in an ellipse is regular and integrable like in a circle, the average ray density does not show a strong dependence on an incident angle.

In order to verify the validity of this idea of nonresonant pumping, we have experimentally investigated the pumping angle dependence in a QDM, which is made of a liquid jet of ethanol doped with Rhodamine 6G dye at a concentration of 10$^{-4}$ M/L. The details on our liquid-jet deformed microcavity, including its fabrication and characterization, are described elsewhere \cite{Yang-RSI06}. The microcavity used in the experiment was best fit by a quadrupole with $\eta$=16\%. The liquid jet apparatus was mounted in the center of a computer-controlled rotational stage so that the pumping angle could be changed at will. For the pump laser, an Ar$^+$ laser operating at $\lambda$=514 nm was used to excite the QDM column from the side. The polarization of the pump laser was parallel to the cavity column. The size parameter $nka$ of our QDM is about 250 at the pump wavelength, where $k=2\pi/\lambda$ the wave vector.

\begin{figure}
\includegraphics[width=3.3in]{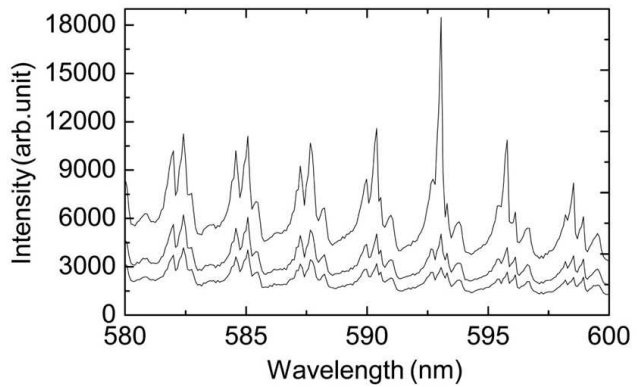}
\caption{Cavity-modified fluorescence and lasing spectra of our QDM for pump power of 22 mW, 33 mW and 57 mW.}
\label{spectrum}
\end{figure}

Figure \ref{spectrum} shows the cavity-modified fluorescence (CMF) and lasing spectra emitted from our QDM for a fixed pumping angle of 56$^\circ$. There exist several groups of cavity modes with well defined free spectral ranges in that spectrum. The detailed study on mode groups in our QDM is presented elsewhere \cite{Lee-PRL06,prep.PRA.Shim05}. As the pumping power increases, a single mode around 593 nm undergoes a laser oscillation. This mode is found to be a scar mode \cite{Lee02} with a radial quantum number or a mode order of $l=1$, corresponding to an eight-fold unstable periodic orbit of ray in the limit of classical chaos \cite{prep.PRA.Shim05}. From both the laser threshold analysis \cite{Lee02,Moon00} and the CMF analysis \cite{Chylek91}, this mode is found to have $Q$ of 5$\times 10^5$. 

\begin{figure}
\includegraphics[width=3.3in]{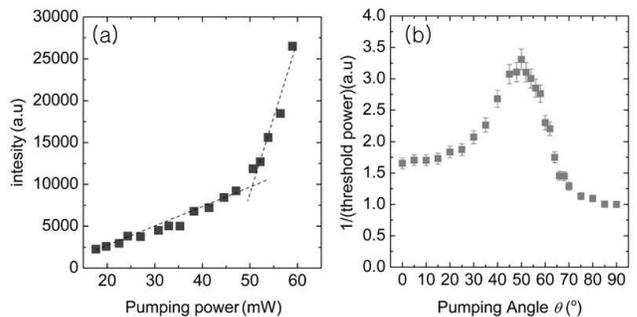}
\caption{(a) Lasing threshold curve for a pumping angle of 50$^o$. (b) Inverse of the threshold pump power vs.\ the pumping angle, normalized with respect to $\theta=\pi/2$.}
\label{efficiencyvsangle}
\end{figure}

By recording the peak height of the $l=1$ mode in Fig.\ \ref{spectrum} as the pumping power is varied, a lasing threshold curve can be obtained. A threshold pumping power is then given by the intersection point of two fitting lines with different slopes as shown in Fig. \ref{efficiencyvsangle}(a). We repeat the same procedure for different pumping angles and finally plot the inverse of the threshold pump power as a function of the pumping angle as shown in Fig.\ \ref{efficiencyvsangle}(b). Since the threshold pump power is inversely proportional to the pumping efficiency, Fig.\ \ref{efficiencyvsangle}(b) can be considered as a plot of the pumping efficiency vs.\ the pumping angle. The maximum pumping efficiency is observed when $\theta \simeq 50^\circ$, at which the pumping efficiency is three times higher than that of $\theta= 90^\circ$, the minimum efficiency. Overall, the observed pumping efficiency curve resembles the prediction by the ray model, but there also exist some differences in details. For example, the step-like sharp feature around $\theta\sim 60^\circ$ seen in Fig.\ \ref{Raymodel}(b) is not present in the observed pumping efficiency curve. In the ray model this feature coincides with the onset of two- and three-bounce ray trajectories inside the cavity as $\theta$ is scanned from 90$^\circ$ to 0$^\circ$.

\begin{figure}
\includegraphics[width=3.3in]{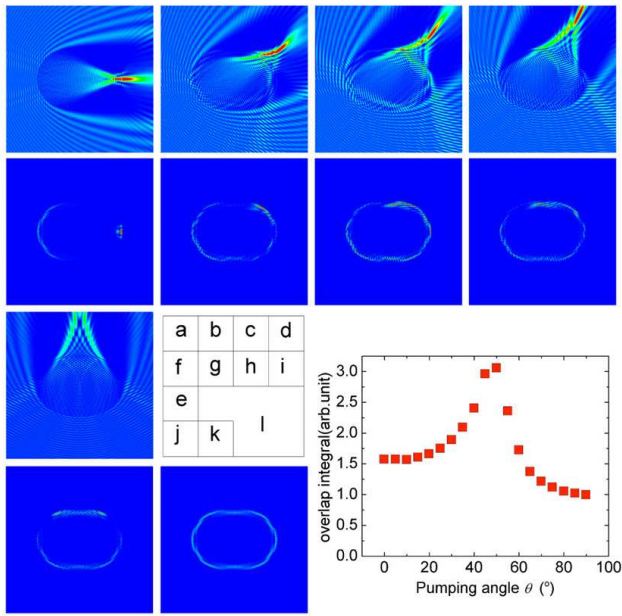}
\caption{Calculated wave distributions when a plane wave satisfying $nka=68.05$ is incident at various pumping angles: (a) $90^\circ$, (b) $60^\circ$, (c) $45^\circ$, (d) $30^\circ$ and (e) $0^\circ$. The pump wavelength is in nonresonant condition. A scar mode of $l=1$ with $nka=96.31$ is shown in (k). In (f)-(j), intensity-intensity overlap with the scar mode is shown just below each wave distribution. (l) Overlap integral as a function of the pumping angle, normalized with respect to $\theta=\pi/2$.}
\label{wanderingtime}
\end{figure}

Since the size parameter of our QDM at the pump wavelength, $nka\sim 250$, is by no means large enough to exclude any wave nature in the buildup of the pump field inside, we also calculated the wave function density for various incident angles. Since this particular problem of plane wave scattering becomes prohibitively massive in numerical computation for $nkr>80$, our calculations were carried out for $nka\sim 70$. The results are shown in Figs.\ \ref{wanderingtime}(a)-(e) for a set of representative pumping angles. When the pump beam is incident at $30^\circ$, $45^\circ$ and $60^\circ$ an enhanced intensity distribution is seen near the cavity boundary due to multiple total internal reflections. 

Since the wave calculation for the pump field is done for $nka$ about three times smaller than the actual experiment, direct comparison of the wave calculation result with the experiment should be done with a care. Nonetheless, we can calculate the overlap integral of the wave function densities in Figs.\ \ref{wanderingtime}(a)-(e) with a scar mode wave function density shown in Fig.\ \ref{wanderingtime}(k), which was calculated by using the boundary element method \cite{Kagami84,Wiersig03,prep.PRA.Shim05} for a size parameter $nka=96.31$, about a half of the experimental size parameter at the lasing wavelength. The overlap integral, proportional to the pumping efficiency, is shown in Fig.\ \ref{wanderingtime}(l). Interestingly, this result is more closer to the experimental data than the result of the ray tracing model is in that the step-like sharp feature is not seen and the value at 0$^\circ$ is about as large as that in the experiment, indicating that the wave nature of the pump field also plays an important role in our pumping scheme. 

In conclusion, we have demonstrated efficient nonresonant optical pumping in a QDM and explained the observed pumping angle dependence in both ray and wave pictures. The present nonresonant pumping technique can be applied to microlasers and filters based on ARC. In a deformation-tunable microcavity, recently demonstrated in Ref.\ \cite{Yang-RSI06}, resonance frequencies of scar modes can be tuned by slightly changing the deformation. In this case, the present technique of nonresonant pumping can readily be adapted to a resonant pumping scheme for development of ultralow threshold scar-mode lasers with high output directionality \cite{Lee02}.

This work was supported by National Research Laboratory Grant and by KRF Grant (2005-070-C00058). JBS and HWL were supported by a Grant from the Ministry of Science and Technology of Korea. SWK was supported by KOSEF Grant (R01-2005-000-10678-0).

\bibliographystyle{prsty}

\end{document}